\documentclass[lettersize,journal]{IEEEtran}

\usepackage{scalerel}
\usepackage{tikz}
\usetikzlibrary{svg.path}

\definecolor{orcidlogocol}{HTML}{A6CE39}
\tikzset{
  orcidlogo/.pic={
    \fill[orcidlogocol] svg{M256,128c0,70.7-57.3,128-128,128C57.3,256,0,198.7,0,128C0,57.3,57.3,0,128,0C198.7,0,256,57.3,256,128z};
    \fill[white] svg{M86.3,186.2H70.9V79.1h15.4v48.4V186.2z}
                 svg{M108.9,79.1h41.6c39.6,0,57,28.3,57,53.6c0,27.5-21.5,53.6-56.8,53.6h-41.8V79.1z M124.3,172.4h24.5c34.9,0,42.9-26.5,42.9-39.7c0-21.5-13.7-39.7-43.7-39.7h-23.7V172.4z}
                 svg{M88.7,56.8c0,5.5-4.5,10.1-10.1,10.1c-5.6,0-10.1-4.6-10.1-10.1c0-5.6,4.5-10.1,10.1-10.1C84.2,46.7,88.7,51.3,88.7,56.8z};
  }
}

\newcommand\orcidicon[1]{\href{https://orcid.org/#1}{\mbox{\scalerel*{
\begin{tikzpicture}[yscale=-1,transform shape]
\pic{orcidlogo};
\end{tikzpicture}
}{|}}}}

\usepackage{amsmath,amsfonts}
\usepackage{algorithmic}
\usepackage{algorithm}
\usepackage{array}
\usepackage[caption=false,font=normalsize,labelfont=sf,textfont=sf]{subfig}
\usepackage{textcomp}
\usepackage{stfloats}
\usepackage{url}
\usepackage{verbatim}
\usepackage{graphicx}
\usepackage{cite}

\usepackage{hyperref} 

\begin{document}

\title{On fusing active and passive acoustic sensing for simultaneous localization and mapping}

\author{Aidan J Bradley\orcidicon{0000-0002-1996-9458}\,,~\IEEEmembership{Graduate Student Member,~IEEE,} and Nicole Abaid\orcidicon{0000-0002-0053-4710},~\IEEEmembership{Member,~IEEE}
\thanks{Manuscript received XXXX; revised XXXX.}
\thanks{Aidan Bradley is with the Engineering Mechanics Program at Virginia Tech, Blacksburg, VA 24061, U.S.A.(e-mail: aidjbradley@vt.edu)}
\thanks{Nicole Abaid is with the Department of Mathematics at Virginia Tech, Blacksburg, VA 24061, U.S.A.(e-mail: nabaid@vt.edu)}}

\markboth{~IEEE Transactions on Signal Processing,~Vol.~XX, Year}%
{Shell \MakeLowercase{\textit{et al.}}: A Sample Article Using IEEEtran.cls for IEEE Journals}

\IEEEpubid{0000--0000/00\$00.00~\copyright~2021 IEEE}

\maketitle

\begin{abstract}
Studies on the social behaviors of bats show that they have the ability to eavesdrop on the signals emitted by conspecifics in their vicinity. They can fuse this ``passive" data with actively collected data from their own signals to get more information about their environment, allowing them to fly and hunt more efficiently and to avoid or cause jamming when competing for prey. Acoustic sensors are capable of similar feats but are generally used in only an active or passive capacity at one time. Is there a benefit to using both active and passive sensing simultaneously in the same array? In this work we define a family of models for active, passive, and fused sensing systems to measure range and bearing data from an environment defined by point-based landmarks. These measurements are used to solve the problem of simultaneous localization and mapping (SLAM) with extended Kalman filter (EKF) and FastSLAM 2.0 approaches. Our results show agreement with previous findings. Specifically, when active sensing is limited to a narrow angular range, fused sensing can perform just as accurately if not better, while also allowing the sensor to perceive more of the surrounding environment. 
\end{abstract}

\begin{IEEEkeywords}
Array Processing, Extended Kalman Filter, FastSLAM 2.0, Multistatic sonar, SLAM.
\end{IEEEkeywords}

\section{Introduction}
\label{s:Intro}
\IEEEPARstart{W}{hether} robotic vehicles using acoustic sensors can improve their simultaneous localization and mapping (SLAM) capabilities by taking advantage of both signals created by their own sensors (i.e. active sensing) and those created by other vehicles (i.e. passive sensing) is an open question in engineering. Research on collective behavior in bats suggests they not only use echolocation but can also eavesdrop on signals from other bats around them. They use this information to both cause and avoid signal jamming from conspecifics \cite{chiu2008batsilence, corcoran2014batsjamming, bates2008jamming} and form temporary mobile networks which improve prey localization \cite{roeleke2022batnetwork}. Bats perform these feats with a relatively simple set of ``sensors" which consist of a single emitter, their larynx, and two receivers, their ears. While a bat's ``hardware" may seem simple enough, the key to the sophistication of their abilities lies in the way they process the signals they create. The ability to adapt their sensing to different environments and prey \cite{fenton1998batcomps, jones2007batcalladapt} and modulate the directionality \cite{arditi2015batmouth}, temporal shape \cite{obrist1995flexible}, and frequency \cite{bates2008jamming} of their calls have inspired the creation of biomimetic sensors that seek to harness these same capabilities \cite{fu2016dynamicbathead, yang2018designbathead, eliakim2018batrobot}. Outside of biomimetics, engineers have been able to match the performance of many of these systems for years using large arrays of transducers and intelligently designed algorithms \cite{van2002optimum}. This work seeks to continue and improve upon previous research and results found in \cite{jahromi2021eavesdropping}, to find if modern sonar -- using techniques inspired by the eavesdropping of bats -- can answer the question of whether fusin active and passive sensing signals may benefit a sonar application.

Just as bats may use active and passive sensing techniques, all sensors for automated systems can generally be partitioned into those two groups as well, each of which come with their own benefits and challenges \cite{christensen2016sensing}. Passive sensors such as cameras and microphones detect signals produced by other sources in the environment, such as reflected light and direct or echoed sound. While they generally do not have an ability to detect distance directly from a single measurement, not having to emit energy into the surrounding environment and having generally longer effective ranges due to a single direction of travel are typically beneficial \cite{urick1983principles}. Conversely, active sensors such as lidar, radar, and active sonar emit energetic signals into the environment and process return data. These returns allow for estimation of range and bearing data and even identification of objects in environments where a camera would be unable to see \cite{belcher2001sonarID}. However, active sensors require more power to create signals and may need extra signal processing to avoid jamming in multiagent systems \cite{garcia2015spreading}. 

\IEEEpubidadjcol

While acoustic sensors are mainly used in underwater applications, due to the ability of sound to travel very long distances compared to the rapid attenuation of radio waves and light \cite{stojanovic2009underwater}, the low cost of acoustic sensors also make them popular for in-air robotic applications and vehicle sensors \cite{chong2015sensor, toa2020ultrasonic}. These sensors are not without their challenges, mainly due to the vast difference in the speed of sound compared to the speed of light. In long distance or fast moving applications, this leads to discrepancies in vehicle position between signal emission and echo reception when using active sonar or being tracked by passive sonar \cite{leonard2012directed}. Some solutions to this issue include incorporating the vehicle dynamics into the estimation using Doppler shifts \cite{jia2020localization} or using synchronization to enable one-way travel-time localization using a beacon and receiver \cite{rypkema2017one}. Further, the longer wavelengths of acoustic signals (as compared to light) \cite{urick1983principles, toa2020ultrasonic} lead to difficulty in detecting objects that are small relative to the given wavelength. In addition, signals that reflect off surfaces at steep angles are reflected specularly with most energy being spread away from the sensor, leading to errors or misses in detection and estimation \cite{brown1985feature, kleeman2016sonar}. Acoustic sensors also tend to be quite directional. It can be difficult to detect signals outside of the main response axis (MRA) when directivity is high, but low directivity makes it difficult to estimate the bearing of a return signal. This can be solved by using mechanically scanned sensors \cite{he2012auv}, but most modern solutions make use of static arrays to improve directivity and well studied electronic steering methods to estimate bearing of echoes \cite{waite2002sonar, van2002optimum}, rather than introduce the complexity of a dynamic sensor.

Inspired by the echolocation and eavesdropping abilities of bats and the capabilities of sonar arrays to act as both active and passive sensors, we use the problem of landmark-based SLAM to investigate possible benefits of fusing these two types of sensing. In short, SLAM deals with a robot building a map of an unknown environment and localizing itself within that environment. Two well studied solutions to this problem are extended Kalman filtering (EKF)-SLAM and FastSLAM 2.0 \cite{thrun2002probabilistic, montemerlo2007fastslam}. While not optimal, the EKF algorithm has been shown to be very effective in solving the SLAM problem \cite{durrant2006simultaneous}. Using EKF also provides a direct link to benchmark results against the previous data found in \cite{jahromi2021eavesdropping}. FastSLAM 2.0 has also been shown to be effective and at times even better than EKF-SLAM \cite{kurt2012comparison}, though it is more complex to implement and much more prone to inconsistency \cite{bailey2006Fastconsistency}. While these algorithms are not the most advanced in terms of accuracy, their long and well researched history is important for analysis of this new technique and informing the future directions of this work.

Previous research \cite{jahromi2021eavesdropping} shows that, when the angle of the active sensor is narrow and measurement noise is low, there is a benefit to being able to take advantage of both active and passive sensing at the same time. Building on previous work, this study introduces physics-based models of active, passive, and fused acoustic sensing that benefit from improved passive landmark initialization and use increased map size. We compare the performance of fused sensing between two well studied SLAM solutions, to analyze differences in regions where fused sensing is effective. The rest of the paper is organized as follows. Section \ref{s:SysModels} describes the models used to simulate the system and obtain the measurement data used in the SLAM algorithms. Section \ref{s:SLAM} briefly presents the SLAM algorithms and goes in depth on landmark initialization. Sections \ref{s:Results} and \ref{s:Discuss} present and discuss results and data interpretation respectively, and Section \ref{s:Conclude} provides concluding remarks.

\begin{figure*}[ht]
    \centering
    \includegraphics{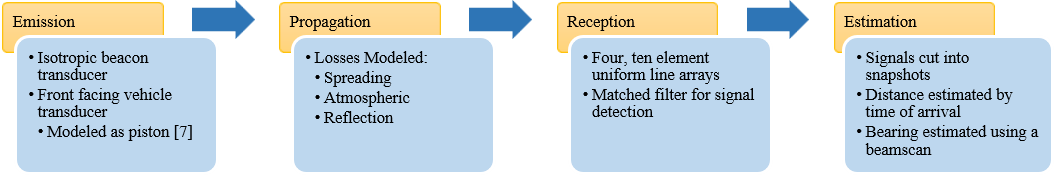}
    \caption{A workflow of how signals are modeled in the system.}
    \label{fig:workflow}
\end{figure*}

\section{System Models}
\label{s:SysModels}
To provide a simulation that is more faithful to real acoustic performance, we chose to model the transducer performance, signal propagation, and range and bearing estimation. In previous work \cite{jahromi2021eavesdropping}, sensing bearings and ranges were set as predefined geometric shapes and the passive sensing range was considered a static radius around the vehicle. While this may capture the main response for cameras and lidar, the performance of a single sonar transducer or, even more so, an array, is defined by a beampattern which is a function of their design and array geometry \cite{van2002optimum}. Further, whether or not a vehicle can sense an echo off a nearby landmark is not only a matter of the vehicle-to-landmark distance and angle, but also a matter of the distance between the landmark and a sound source in the environment that may enconify it. A workflow of the simulation can be seen in Fig. \ref{fig:workflow} and the models are discussed in detail below.

\subsection{Emitter Modeling}
\label{ss:Emitter}
The receiving arrays and emitters of the system are modeled using MATLAB's Phased Array System Toolbox \cite{phasedArray}.  Simulating the behavior of the transducer system was of interest as, unlike in \cite{jahromi2021eavesdropping}, sonar transducers do not have discrete cutoffs in bearing and range where they can no longer operate and the noise of measurements cannot be assumed to be constant throughout the effective range of the sensor. Realistically, these transducers have beam patterns with a main response axis (MRA) that can be physically or electronically steered for estimating range and bearing measurements, and multiple side lobes that leak energy when emitting or can confound bearing estimations when receiving signals in the presence of other noise \cite{van2002optimum}.

We define two emitters that ensonify the environment, one on a static beacon and the other on a mobile vehicle. The two emitters are each initially modeled as isotropic projectors with frequencies of 30 kHz and 35 kHz respectively, so that the signals can be distinguished from one another. While the beacon emission beampattern remains isotropic for all simulations, the vehicle emitter is further backbaffled, meaning it only emits in a 180$^\circ$ arc towards the vehicle's direction of motion, and its directionality is finally adjusted using the piston model from \cite{arditi2015batmouth}
\begin{equation}
    \label{eqn:pistonModel}
    p(\phi_{i_{\mathrm{true}}},f) = \frac{\mathrm{abs}[2J_1(2\pi \frac{f}{c})a\sin(\phi_{i_{\mathrm{true}}})]}{(2\pi \frac{f}{c})a\sin\phi_{i_{\mathrm{true}}}}
\end{equation}

\begin{figure}
\centering
\includegraphics{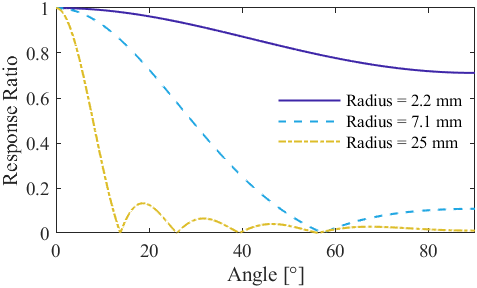}
\caption{Half of the beampattern created by piston model (\ref{eqn:pistonModel}) where $0^{\circ}$ is the MRA of the emitter. Response is given as a linear ratio to improve readability.}
\label{fig:pistonMode}
\end{figure}

Here, $p(\phi_{i_{true}},f)$ is the pressure ratio between the MRA of the vehicle emitter and the actual bearing angle between the vehicle's direction of travel and a landmark in the domain which is ensonified at a given frequency, where $\phi_{i_{true}}$ is the actual angle of the landmark relative to the vehicle. While $f$ is the frequency of the emitter signal, $c$ is the speed of sound in air at standard temperature and pressure, $J_1$ is the first order Bessel function of the first kind, and $a$ is the radius of the emitter (i.e. the piston). The larger $a$ is, the narrower the MRA of the emitter becomes and more side lobes are introduced to the beam pattern. The effect of the emitter radius on the beampattern of the emitted signal can be seen in Fig. \ref{fig:pistonMode}. In this work, the determining factor of the emitter opening angle is considered to be the half-power beamwidth (HPBW) \cite{van2002optimum}, whichh is the beamwidth at which half the maximum power (0.5 in Fig. \ref{fig:pistonMode}) is achieved.

\subsection{Signal Propagation}
\label{ss:2SigProp}
The acoustic signals are modeled as plane waves that travel along rays directly from an emitter to a receiver \cite{urick1983principles,landau2013fluid}. While only signals from the vehicle's emitter experience loss from the piston model (\ref{eqn:pistonModel}), all emitted signals also experience spreading, atmospheric, and reflection losses. Spreading losses in decibels are calculated using the spherical spreading equation:
\begin{equation}
    \label{eqn:sphSpreading}
    \beta = r_{\mathrm{factor}}20\log_{10}\frac{d}{r_{\mathrm{factor}}},
\end{equation}
where $d$ is the distance the signal has traveled from the source and $r_\mathrm{factor}$ takes a value of $1$ if the signal only travels in one direction, or $2$ if the signal has two way travel back to the source, as it does in the case of active sensing. Atmospheric losses are calculated using a formula from \cite{blackstock2001fundamentals}, which in the case of standard temperature and atmospheric pressure (STP) has the form:
\begin{multline}
\label{eqn:atmosLoss}
\text{$\alpha = 8.686f^2\bigg(1.84\times10^{-11} +$}\\
\text{$6.1424\times10^{-6}\frac{f_{rO}}{f_{rO}^2+f^2} +$}\\
\text{$1.5552\times10^{-6}\frac{f_{rN}}{f_{rN}^2+f^2}\bigg)$},
\end{multline}
where $\alpha$ is the loss factor, $f$ is the frequency of the emitted signal, $f_{rN}$ is the relaxation frequency of nitrogen, and $f_{rO}$ is the relaxation frequency of oxygen at STP. Finally, the landmarks in the environment are modeled as small spheres with a target strength calculated using:

\begin{multline}
    \label{eqn:tgtLoss}
    \text{$I(\gamma_{LM}) = \frac{r_{LM}}{2}\sin(\gamma_{LM}) +$}\\
    \text{$\frac{1}{2\pi k}\cot\left(\frac{\gamma_{LM}}{2}\right)^2\sin\left(r_{LM}k\sin\left(\gamma_{LM}\right)\right)$},
\end{multline}

\noindent where $\gamma_{LM}$ is the angle of the reflection with $180^{\circ}$ being directly back towards the direction of the emitter, $I(\gamma_{LM})$ is the intensity of the reflected signal, $r_{LM}$ is the radius of the target, and $k = 2\pi f/c$ is the wavenumber of the reflected signal. The intensity is converted to decibels using $10\log_{10}(I(\gamma_{LM})/I(180^{\circ}))$. It is of note that the Doppler effect of the moving vehicle on the frequencies it receives are also modeled, but the movement of the vehicle is not fast enough for it to make an appreciable difference.

\subsection{Receiving Array}
\label{ss:2receiver}
Receiving arrays are modeled as 4 sets of 10 element uniform line arrays (ULA) arranged as a square with the ``front" array orthogonal to the vehicle's direction of motion. The number of elements for each array was chosen to balance directionality, and therefore bearing measurement accuracy, and what could feasibly fit on a small size robot. Each element of the array was modeled as a backbaffled isotropic receiver and their spacing was designed to be half the wavelength of the frequency of the highest signal to be measured, which is the beacon signal in our case. This spacing was chosen to avoid the introduction of grating lobes when electronically steering the MRA of the arrays to their limits of $\pm 80^{\circ}$ \cite{van2002optimum}

\subsection{Signal Detection and Processing}
\label{ss:2SigProcess}
After all losses are applied to each echo, the time series of data is passed to a receiver preamp, an object in the phased array toolbox, which adds a gain and noise to the data. Each of the four ULAs has its own time series the length of the timestep, that is further partitioned into individual snapshots of data using a matched filter, to find sections of time where echoes are present. Each snapshot may contain a single signal, or multiple signals if the echoes from different landmarks happen to overlap in time. Using these snapshots provides numerous benefits to the analysis of the signal data: it enables the system to more easily calculate the spectrum of each echo to determine whether the signal came from the vehicle itself (an ``active" echo) or the beacon (a ``passive" echo), the flight time for actively sensed echoes can be more easily calculated using a phase shift beamformer \cite{van2002optimum}, and obfuscation of signals arriving at different times but from the same bearing is more easily avoided.

After snapshots are selected for each ULA, the spectrum of each is analyzed to find whether it contains only active or passive echoes or both. If the snapshot is determined to include active echoes then both range and bearing measurements must be estimated. This is done by first using a beamscan algorithm which electronically forms a conventional beam and scans it across bearings of interest \cite{van2002optimum}. Each bearing scanned produces a single power value and the peaks of these powers suggests a possible arrival direction of an echo. Each peak over a set threshold is selected as a possible bearing measurement and a beamformer is used to calculate the time series of the signal in these selected directions. The beamformer amplifies signals coming from the selected direction while reducing signals in other directions, dependent on the beampattern of the array. If an echo is in fact coming from the chosen direction, then its time of arrival can be determined, which means both range and bearing measurements of the echo are available. 

In the case of a passive only simulation, only a bearing measurement is of interest as we do not assume the timing of the beacon emission is known. It has been shown that it is possible for the beacon and vehicle to be synchronized so the vehicle can know when the beacon emits  \cite{rypkema2017one}, but these techniques are outside the scope of this paper. If both active and passive echoes are available, the same algorithms are run and the simulated array beampattern is changed based on the signals present in a given snapshot, as the beampattern is a function of the array geometry and signal frequency \cite{van2002optimum}.

It must be noted that a beamscan algorithm was chosen over numerous optimal and adaptive beamforming algorithms such as those found in \cite{van2002optimum, huang2019new, du2009review}, many of which are available in the phased array MATLAB toolbox, as we were interested in investigating the limits of the SLAM algorithms with regard to measurement accuracy. Using the most straightforward implementation of these systems and not trying to create the most accurate measurement system possible for these situations can give us a better idea of baseline performance.

Based on the above, there are multiple measurements corresponding to each array's local coordinate system spanning from $\pm160^{\circ}$. These are rotated to the vehicle's local coordinate system and repeated measurements are averaged. These final measurements, which span a full $360^{\circ}$, are passed to the SLAM algorithms to be used for estimation.

\subsection{Estimation Models}
\label{ss:2EstModel}
The motion model for this system, presented in \cite{bailey2006EKFconsistency}, is described as ``the kinematic model for the trajectory of the front wheel of a bicycle subject to rolling motion constraints (i.e., assuming zero wheel slip)", with the form:
\begin{equation}
    \label{eqn:motionModel}
    \mathbf{x}_k \hspace{-3pt} = \textrm{f}(\mathbf{x}_{k-1}, \mathbf{u}_k) = 
    \begin{bmatrix}
        x_{v_{k-1}} + V_k\Delta_t\cos(\theta_{v_{k-1}}+\gamma_k) \\
        y_{v_{k-1}} + V_k\Delta_t\sin(\theta_{v_{k-1}}+\gamma_k) \\
        \theta_{v_{k-1}}+\frac{V_k\Delta_t}{B}\sin(\gamma_k)
    \end{bmatrix},
\end{equation}

\noindent where, over time $\Delta_t = k - (k-1)$ the control values $u_k = [V_k,\gamma_k]^T$, speed and steering angle respectively, are kept constant and $B$ is the wheelbase between the front and rear axles. Additive noise $\nu_k = [\nu_{V_k}, \nu_{\gamma_k}]^T$ is sampled from the Gaussian distribution $\mathcal{N}(0,Q)$, where the process noise covariance matrix has the form $Q = \mathrm{diag}[\sigma_V, \sigma_\gamma]$ and is added directly to $\mathbf{u}_k$. Here, $\mathrm{diag}[\cdot]$ denotes an appropriately-sized diagonal matrix with the argument on the diagonal.

We use two vehicle-to-landmark measurement models to calculate predicted measurements, which are dependent on whether the signal being received is considered an active or passive measurement. Active measurements in two dimensions provide range and bearing data from the currently estimated vehicle state $\hat{\mathbf{x}} = [\hat{x}_v,\hat{y_v},\theta_v]$ to the respective landmark $\hat{\mathbf{m}}_i = [\hat{x}_i, \hat{y}_i]^T$ and the measurement model for a landmark $i$ is given by:
\begin{equation}
    \label{eqn:actModel}
    \mathbf{\bar{z}}_{i}^{act} = h_i^{act}(\hat{\mathbf{x}},\hat{\mathbf{m_i}})=
    \begin{bmatrix}
        \sqrt{(\hat{x}_i-\hat{x}_v)^2 + (\hat{y}_i-\hat{y}_v)^2} \\
        \tan^{-1}(\frac{\hat{y}_i-\hat{y}_v}{\hat{x}_i-\hat{x}_v}) - \hat{\theta}_v\\
        \hat{c}_{i}
    \end{bmatrix}+\omega_k,
\end{equation}
\noindent where a bar over a value indicates the predicted value. Passive measurements in two dimensions only provide bearing data from the current position to $\hat{\mathbf{m}}_i$ and therefore have the form:
\begin{equation}
    \label{eqn:pasModel}
    \mathbf{\bar{z}}_i^{pas} = h_i^{pas}(\hat{\mathbf{x}},\hat{\mathbf{m_i}})=
    \begin{bmatrix}
        \tan^{-1}(\frac{\hat{y}_i-\hat{y}_v}{\hat{x}_i-\hat{x}_v}) - \hat{\theta}_v \\
        \hat{c}_{i}
    \end{bmatrix}+\omega_k.
\end{equation}

The general measurement vector of a landmark $i$ is of the form $\mathbf{z}_i=[d_i,\phi_i,c_i]^\textrm{T} $, with the vehicle-to-landmark range and bearing measurement being $d_i$ and $\phi_i$ respectively and the measured landmark identification being $c_i$. Additive measurement noise $\omega = [\omega_d, \omega_{\phi}, \omega_c]^T$ is sampled from the Gaussian distribution $\mathcal{N}(0,R)$, where the measurement noise covariance matrix has the form $R=\mathrm{diag}[\sigma_d,\sigma_\phi, \sigma_c]$. Readers should note, for passive measurements, the range measurement and noise is not considered. For the rest of the paper unless explicitly stated, $\bar{\mathrm{z}}$ will represent predicted measurements for both active and passive cases.

\section{SLAM Algorithms}
\label{s:SLAM}
We implemented two landmark-based SLAM algorithms, EKF-SLAM and FastSLAM 2.0, to analyze the measurement data simulated from the model above. Both of these solutions were chosen due to their maturity, while also allowing us to compare the performance of two solutions that have different strengths and weaknesses \cite{kurt2012comparison, brooks2009hybridslam}. 

\subsection{EKF-SLAM}
EKF-SLAM is one of the most common and mature algorithms used to solve SLAM problems, with countless summaries and explanations to be found in literature, see \cite{thrun2002probabilistic, durrant2006simultaneous,khairuddin2015review}.  The state of the estimated map is a joint random state-vector, $\mathbf{X}$, composed of the estimated vehicle state and landmark states
\begin{equation}
\label{eqn:ekfMapState}
    \hat{\mathbf{X}} = [\hat{x}_{v}, \hat{y}_{v}, \hat{\theta}_{v}, \hat{x}_1, \hat{y}_1, \hat{c}_1, ..., \hat{x}_N, \hat{y}_N, \hat{c}_N]^T = 
    \begin{bmatrix}
        \hat{\mathbf{x}} \\ 
        \hat{\mathbf{m}} 
    \end{bmatrix}
\end{equation}

\noindent The vehicle's 2-D planar coordinates and global bearing $\mathbf{\hat{x}}$ at timestep $k$ are given as $\hat{x}_{v}, \hat{y}_{v}$, and $\hat{\theta}_{v}$ respectively, while the stationary map parameters, $\hat{\mathbf{m}} = [\hat{x}_1, \hat{y}_1, \hat{c}_1,..., \hat{x}_N, \hat{y}_N, \hat{c}_N]^T$, are a given landmark's 2-D coordinates and landmark number. The landmark numbers $\hat{c}_i, \, i=1,\ldots,N$ are assigned in order of their initialization on the map. To complete the Gaussian \textit{a posteriori} probability density approximation of $\mathbf{\hat{X}}$, a covariance matrix is also defined as:
\begin{equation}
    \label{eqn:aPostCov}
    \textsc{P} = \begin{bmatrix}
                    P_\textbf{xx}   &   P_\textbf{xm}\\
                    P_\textbf{mx}   &   P_\textbf{mm}
                \end{bmatrix}.
\end{equation}

\noindent The recursive procedure of the EKF algorithm can be summarized by two general steps. The first step is prediction, where the zero noise motion model of the system (\ref{eqn:motionModel}) is used to propagate the current vehicle state and the calculation of the \textit{a priori} covariance matrix is:
\begin{equation}
    \label{eqn:aPriorCov}
    \begin{matrix}
        \text{$\bar{P}_\textbf{xx} =FP_\textbf{xx}F^\textsc{T} + UQU^\textsc{T}$}\\
        \text{$\bar{P}_\textbf{mx}^\textsc{T} = \bar{P}_\textbf{xm} = FP_\textbf{xm}$}
    \end{matrix},
\end{equation}

\noindent with $F = \partial\textrm{f}/\partial \textbf{x}|_{(\hat{\textbf{x}}_k, u_k)}$ and $U = \partial\textrm{f}/\partial u|_{(\hat{\textbf{x}}_k,u_k)}$. The second step is a correction conditioned on the landmark measurements at the current time step, which can be completed as a batch or individually. For a given landmark $i$, the calculations are as follows:
\begin{equation}
    \label{eqn:EKFCorrect}
    \begin{matrix}
        \text{$K_i =\bar{P}H_i^\textrm{T}(H_i\bar{P}H_i^\textrm{T} + R)^{-1}$} \\
        \text{$\hat{P} = (I - K_i \ H_i) \bar{P}$} \\
        \text{$\hat{\mathbf{X}} = \bar{\mathbf{X}} + K_i(z_i - \bar{z}_i)$}
    \end{matrix},
\end{equation}

\noindent where $K_i$ is the Kalman gain and $H_i = \partial h/\partial\mathbf{X}|_{(\bar{\mathbf{X}})}$. Full descriptions and derivations of the EKF-SLAM algorithm can be found in sources such as \cite{thrun2002probabilistic, durrant2006simultaneous, yan2009review}.

\subsection{FASTSLAM 2.0}
FastSLAM 2.0 is almost as well studied as EKF-SLAM \cite{thrun2002probabilistic, khairuddin2015review} but is built upon a different set of foundational algorithms and assumptions \cite{montemerlo2003fastslam, montemerlo2007fastslam}. For this algorithm, the estimated map is a set of $[\mathbf{Y}^{[1]},\mathbf{Y}^{[j]},...\mathbf{Y}^{[N_{\mathrm{part}}]}]$ particles where each particle has the form:
\begin{equation}
\label{eqn:fastMapState}
    \mathbf{Y}^{[j]}=\langle\ \hat{\mathbf{x}}^{[j]}, \{\hat{\mathbf{m}}^{[j]}_1, \hat{P}^{[j]}_1\},...,\{\hat{\mathbf{m}}^{[j]}_N, \hat{P}^{[j]}_N\}\rangle.
\end{equation}

\noindent Here, the vehicle's 2-D coordinates and global bearing are given as the first element, in the same form as in the EKF-SLAM case. Each successive element is a given landmark's 2-D coordinates $\hat{\mathbf{m}}^{[j]}_i = (\hat{x}^{[j]}_i, \hat{y}^{[j]}_i)$ and covariance matrix $P^{[j]}_i$, in order of initialization on the map. FastSLAM 2.0 also follows a recursive prediction and correction procedure with the addition of a resampling step.

During the prediction step, the vehicle state estimate is sampled from the \textit{a posteriori} distribution:
\begin{equation}
    \label{eqn:fastPrior}
    \hat{\mathbf{x}}^{[j]}_k \sim p(\hat{\mathbf{x}}_k|\hat{\mathbf{x}}_{k-1}^{[j]},u_k,\mathbf{z}_k).
\end{equation}

\noindent Due to the inclusion of the measurement vector $\mathbf{z}_k$, the derivation of this proposal distribution is rather complex so only a brief summary will be given here, while a full derivation can be found in \cite{thrun2002probabilistic}. The proposal distribution is Gaussian and is calculated for particle $[j]$ and landmark $i$ as:
\begin{equation}
    \label{eqn:fastPredict}
    \begin{matrix}
        \text{$R_i = R + H_{m,i}P^{[j]}_iH_{m,i}^\textrm{T}$} \\
        \text{$\hat{P}_\mathbf{x_v}^{[j]} = [H_\mathbf{x_v}^\textrm{T}R_i^{-1}H_\mathbf{x_v} + Q^{-1}]^{-1}$} \\
        \text{$\mathbf{X}^{[j]}_\mathbf{x_v} = \hat{P}_\mathbf{x_v}^{[j]}H_\mathbf{x_v}^\textrm{T}R_j^{-1} (z_k - \bar{z}_k) + \hat{\mathbf{x}}^{[j]}$}
    \end{matrix}.
\end{equation}

\noindent Here, $R_i$ is the measurement information matrix and the Jacobians $H_{m,i} = \partial h/\partial(\hat{x}_i, \hat{y}_i)|_{\hat{x}^{[j]}_i, \hat{y}^{[j]}_i}$ and $H_\mathbf{x_v} =  \partial h/\partial\mathbf{x}|_{(\hat{\mathbf{x}})}$ are with respect to the landmark and vehicle states respectively. Once all measurements have been incorporated, the new vehicle state estimation for particle $[j]$ is sampled from proposal distribution, $\hat{\mathbf{x}}^{[j]} = \mathcal{N}(\mathbf{X}^{[j]}_\mathbf{x_v}, \hat{P}_\mathbf{x_v}^{[j]})$.

Now, each measurement is incorporated again for the correction step. Landmarks that have been initialized in the map and have been measured this time step are corrected with their own EKF following the calculations outlined above (\ref{eqn:EKFCorrect}). All other initialized landmarks are updated with a copy of their mean and covariance from the previous timestep.

The final step of FastSLAM 2.0 is to resample the particles. If resampling were to be ignored, the filter would eventually tend towards only one particle having substantial weight nullifying the usefulness of the filter. Conversely, every resample leads to entire uncertainty histories and map estimations being erased \cite{bar2004estimation}. In this work the particles are resampled using stratified resampling \cite{douc2005comparison}, once the number of effective particles is below a threshold defined as:

\begin{equation}
    \label{eqn:resampleThresh}
    N_\mathrm{eff}=1/\sum w^2 < \mathrm{Threshold},
\end{equation}

\noindent where $w$ is the weight of each particle and $N_{\mathrm{eff}}$ describes the variance of the particle weights. 

Two key details of SLAM which have not yet been discussed are landmark initialization and data association. For this work, we assumed known data association of measurements. While there are numerous ways to relax this assumption \cite{thrun2002probabilistic, costa2004bearing, fortmann1980multi}, it simplifies our abilities to compare the different sensing approaches and SLAM algorithms. Landmark initialization for active, passive, and fused measurements is discussed in the next section.

\subsection{Landmark Initialization}
\label{ss:LMInit}
While using active sensing, each landmark measurement provides both range and bearing data, so (\ref{eqn:actModel}) can be solved with a single measurement. This means, when a new landmark $i$ is measured, it can be initialized into the map for both EKF-SLAM (\ref{eqn:ekfMapState}) and FastSLAM 2.0 (\ref{eqn:fastMapState}) by its position relative to the current vehicle state using:
\begin{equation}
    \label{eqn:activeInit}
    \begin{bmatrix}
        \hat{x}_i\\
        \hat{y}_i\\
        \hat{c}_i
    \end{bmatrix} = 
    \begin{bmatrix}
        \hat{x}_v\\
        \hat{y}_v\\
        0
    \end{bmatrix} +
    \begin{bmatrix}
        d_i \cos(\phi_i + \hat{\theta}_v)\\
        d_i \sin(\phi_i + \hat{\theta}_v)\\
        c_i
    \end{bmatrix}.
\end{equation}

Here, $i$ is the landmark's identifier which starts at 1 and is incremented with each new landmark seen, and $\hat{\cdot}$ is the estimated value of a state. During bistatic or passive sensing, the ability to measure the distance of an echo is lost, due to ignorance of the original emission time. This lack of knowledge means a single measurement no longer provides enough information to completely initialize a landmark's position. There are three main groups of solutions to this problem: delayed, undelayed, and concurrent. Delayed solutions use multiple measurements and separate temporary maps and filters to estimate the range of measured landmarks, until their covariance is low enough to be fully initialized in the main map \cite{munguia2008delayed, bailey2003constrained, costa2004bearing}. Weaknesses of these solutions include a sensitivity to low parallax measurements and divergence if landmarks are initialized too quickly. To avoid building extra uncertainty while waiting for landmarks to initialize, undelayed solutions use techniques such as inverse depth paramaterization \cite{civera2008inverse}, multiple hypotheses \cite{kwok2004efficient}, or  take advantage of the Gaussian definition of landmark positions and initialize the landmark as a conic ray \cite{sola2005undelayed}. These solutions avoid many of the negatives of delayed initialization but can lead to multiplicative or additive increases in map size. A concurrent solution, such as derived in \cite{munguia2010concurrent}, uses two kinds of representations for whether a landmark is in the delayed or undelayed stage of estimation, but comes with its own complexities in system implementation.

As our map is quite large in size for our vehicle speed and sensor accuracy and the number of landmarks is rather sparse, we have decided to implement the undelayed initialization technique used in \cite{sola2005undelayed}. This solution takes advantage of the Gaussian foundation of the EKF and acts as an approximation to a Gaussian Sum Filter (GSF) \cite{alspach1972GSF}, which only additively grows the map size depending on a set of design assumptions. By defining a minimum and maximum sensor range $[s_{min},s_{max}]$ and the conditions $s_1 - \sigma_1 = s_{min}$ and $s_{N_g} + \sigma_{N_g} >= s_{max}$, an initial landmark range hypothesis and total number of hypotheses can be defined by:
\begin{equation}
    \label{eqn:passInit}
    \begin{matrix}
        \text{$s_1 = (1-\alpha)^{-1}s_{min}$}\\
        \text{$\sigma_1$} = \alpha s_1\\
        \text{$N_g = 1 + \mathrm{ceil} \left[ \log_\beta (\frac{1-\alpha}{1+\alpha}\frac{s_{max}}{s_{min}}) \right]$}
    \end{matrix}.
\end{equation}
Here, $s_1$ and $\sigma_1$ are the mean and standard deviation of the first range hypothesis of a new landmark and $\alpha = \sigma/s$ is a constant which should be kept below $30\%$ \cite{peach1995bearings}. $N_g$ defines the total number of Gaussians that will make up the range estimate with $\beta$ controlling the  distance between each successive mean. Each successive distribution hypothesis is calculated by looping through $s_j = \beta s_{j-1}$ and $\sigma_j = \alpha s_{j-1}$ until $j=N_g$. After all hypotheses are calculated, the final result is a conic ray that points in the direction of the bearing estimate, the map vector and covariance matrix grow by $N_g$ rows and columns, and the landmark becomes partially initialized.

To mitigate inconsistency of the filter over repeated measurements and decide which hypotheses to eventually prune, each hypothesis in the ray is initially weighted with a normalized Aggregated Likelihood (AL).
\begin{equation}
    \label{eqn:aggregateLike}
    \Lambda = [\Lambda_1, ..., \Lambda_{N_g}], \Lambda_j = 1/N_g
\end{equation}
\noindent For each new measurement of a landmark, the weight of a hypothesis $j$ is updated and normalized using:
\begin{equation}
    \label{eqn:weightUpdate}
    \begin{matrix}
        \text{$\lambda_j = \exp(-0.5(z_j - \bar{z}_j) S_j^{-1} (z_j - \bar{z}_j)^\textrm{T}/\sqrt{2\pi |S_j|}$}\\
        \text{$\rho_j = (\lambda_j)^n/ \sum_{i=1}^N (\lambda_j)^n$}
    \end{matrix},
\end{equation}
\noindent where $\lambda_j$ is the likelihood of a hypothesis $j$ given a measurement $z_j$, $\rho_j$ is the normalized weight, $S_j = H_i\bar{P}H_i^\textrm{T} + R $, and $n$ is a measure of how much to weight more likely hypotheses of less likely ones. A method the authors of \cite{sola2005undelayed} call Federated Information Sharing is used to update the map and mitigate inconsistency of the filter. This method is derived from the Federated Filter which applies the Principle of Measurement Reproduction \cite{foxlin2002generalized, tupysev1998generalized}, which says the the correction of a random variable by a set of measurement tuples $\{z;R_j\}$ is equivalent to the unique correction by $\{z;R\}$ if
\begin{equation}
    \label{eqn:PMR}
    R^{-1} = \sum R_j^{-1}.
\end{equation}
\noindent By dividing the measurement covariance matrix $R$ by the weight of each hypothesis $\rho_j$ when it is time to run the correction step of the map, the magnitude of the change in the uncertainty covariance of the vehicle and current landmark states due to less likely hypotheses is reduced. 

To prune the least likely estimates, a threshold is defined by the current number of hypotheses $N$ and a variable $\tau$ that is analogous to the probability that a likely estimate will be pruned. After the AL of each hypothesis is updated using $\Lambda^+_j = \Lambda_j\lambda_j$, the new value is compared to the threshold and is discarded if it is too small, that is, if,
\begin{equation}
    \label{eqn:threshold}
    \Lambda_j < \tau/N.
\end{equation}
\noindent After all but one of the hypotheses is pruned, the landmark is characterized as fully initialized.

As each landmark is defined by a Gaussian distribution, we were able to adapt the above technique for FastSLAM 2.0. The steps of the initialization and map updates stay the same as before, but each particle now has $N_g$ estimates for each new landmark that must be updated and pruned using the same steps as described above. With the increase in new landmark hypotheses, each particle now has the form:
\begin{equation}
\label{eqn:fastPassive}
\begin{matrix}
    \text{$\mathbf{Y}^{[j]}=\langle\ \mathbf{x}^{[j]}, \{[\hat{\mathbf{m}}^{[j]}_{1,1},...,\hat{\mathbf{m}}^{[j]}_{1,N_g}]^\textrm{T}, [P^{[j]}_{1,1},...,P^{[j]}_{1,N_g}]\},...$}\\
    \text{$\{[\hat{\mathbf{m}}^{[j]}_{N,1},...,\hat{\mathbf{m}}^{[j]}_{N,N_g}]^\textrm{T}, [P^{[j]}_{N,1},...,P^{[j]}_{N,N_g}]\}\rangle$}
\end{matrix}.
\end{equation}

To the best of our knowledge, this is the first time this ray initialization technique has been used in FastSLAM 2.0. Hence, an optimal solution of how to handle particles having different numbers of estimates for the same landmark has not been investigated yet. We choose to allow the particles to behave independently for the entire simulation, meaning each particle that has more than one state estimate for a landmark goes through the pruning steps of the algorithm until it has collapsed to a single estimate or an active measurement is taken. The effectiveness of retaining all of this information, compared to dropping unlikely hypotheses after a certain number of particles have pruned them, is still an open question.

During simulation using the fused sensing model, both techniques discussed above are used. If a landmark returns both an active and passive signal, the measured bearings are averaged so that no information is fully lost. For a landmark that has been fully initialized as a single Gaussian or set of particles, active and passive measurements are directly applied to the landmark estimate according to their respective measurement model (\ref{eqn:actModel}),(\ref{eqn:pasModel}). If a landmark is actively measured while still partially realized, all but the heaviest weighted hypothesis is kept, the measured range is taken as the true range, and the bearing and correspondence are updated as usual.

\section{Simulation Setup}
\label{s:SimSetup}
\begin{figure}
    \centering
    \includegraphics[width=2.5in, height= 2.5in]{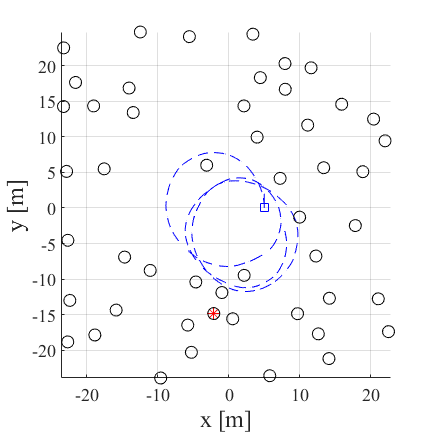}
    \caption{A snapshot of a sample simulation with randomly placed landmarks shown as black circles. The vehicle's current position and its true path are illustrated by the blue square and dashed line, while the beacon, which is also considered a landmark, is illustrated by the red star.}
    \label{fig:exampleMap}
\end{figure}

Simulations are run for the two SLAM algorithms (EKF and FastSLAM 2.0), three sensing strategies (active, passive, and fused), and 12 emitter HPBWs distributed logarithmically in the range $[180^\circ,11^\circ]$ (or $[2.5,25]$ mm according to \ref{eqn:pistonModel}) for active and fused sensing. These HPBWs do not apply to the passive sensing strategy as it uses no directional emitter operated by the vehicle. A logarithmic distribution of beamwidths was chosen as opposed to a linear distribution as in \cite{jahromi2021eavesdropping}, as this previous work suggested the fused sensing technique was most effective at narrow opening angles. A total of $N_{MC} = 115$ Monte-Carlo iterations were performed for each combination of algorithm, sensing strategy, and HPBW (i.e., experimental combinations).

As we include no object avoidance algorithm, the path of the vehicle is first simulated according to (\ref{eqn:motionModel}) with process noise covariance $Q = \mathrm{diag}([0.1^2, 0.03^2])$ and a timestep of $\Delta t = 0.125$ s and the beacon is placed at a random bearing 15 m from the center of the map, which does not interfere with the path of the vehicle. Around these paths, a set of $N = 50$ point landmarks is randomly distributed with a minimum spacing of at least 3 m and a maximum range of 25 m from the center of the map. A minimum spacing of $0.5$ m between each landmark, the beacon, and path of the vehicle is imposed to avoid the system echoes returning before the emitter finishes emitting the signal, a scenario that would be avoided using vehicle control in practical applications. A map is randomly generated for each Monte-Carlo iteration and shared across all experimental combinations. An example of a randomly generated map can be seen in Fig. (\ref{fig:exampleMap}). When the simulation begins, the vehicle is propagated to its next true state $\mathbf{x}$, and after $4$ timesteps or $0.5$ s, both the vehicle and beacon emit and the simulation of signal propagation and measurement estimation described in Section \ref{s:SysModels} begins. The total length of simulation is $187.5$ s, the length of time it would take the vehicle to complete around 3 circles in the map in the absence of process noise.

After the simulation of the vehicle and its measurement are complete, this data is fed to the SLAM algorithms. Both algorithms were tuned with a measurement noise covariance of $R = \mathrm{diag}[0.2^2, 0.15^2, 1^2]$ and given an initial state matching the vehicle's initial state and and initial error covariance of $P_0 = \mathrm{diag}[0.05^2,0.05^2,0.0436^2]$. Each FastSLAM 2.0 simulation was run with 100 particles which were resampled once the number of effective particles fell below $75 \%$ (\ref{eqn:resampleThresh}).

\begin{table*}
\begin{center}
\caption{Simulation Parameters}
\label{tab1}
\begin{tabular}{| c | c | c |}
\hline
\textbf{Variable} & \textbf{Symbol} & \textbf{Value}\\
\hline
Control timestep& $\Delta t$ & $0.125$ s\\
\hline
Measurement timestep& -& $0.5$ s\\
\hline
Number of timesteps& $N_T$& $1500$\\
\hline
Length of simulation& $T$& $187.5$ s\\
\hline
Number of landmarks& $N$& 50\\
\hline
Number of iterations& $N_{MC}$& 115\\
\hline
Vehicle speed& $V$& 0.75 m/s\\ 
\hline
Turning angle& $\gamma$& 0.027 rad\\
\hline
Vehicle wheel base& -& $0.2$ m\\
\hline
std. of vehicle speed& $\sigma_V$& $0.1$ m\\
\hline
std. of turning angle& $\sigma_\gamma$& $0.03$ rad\\
\hline
Emitter HPBW& -& logspace($[11,180]^\circ$)\\
\hline
std. of range measurement noise& $\sigma_d$& 0.2 m\\
\hline
std. of bearing measurement noise& $\sigma_\phi$& 0.15 rad \\
\hline
std. of data association& $\sigma_c$& 1\\ 
\hline
Simulated array elements& -& 10\\
\hline
Array sampling frequency& -& 200 kHz\\
\hline
FastSLAM particles& -& 100\\
\hline
Resampling threshold& -& 75\\
\hline
Initial state covariance& $P_0$& $\mathrm{diag}[0.05^2, 0.05^2, 0.0436^2]$\\
\hline
Passive init. range covariance factor& $\alpha$& 0.3\\
\hline
Passive init. ray density& $\beta$& 3\\
\hline
Range bounds& $[s_{min}, s_{max}]$& $[0.5, 20]$ m\\
\hline
Number Gaussian hypotheses& $N_g$& 4\\
\hline

\end{tabular}
\end{center}
\end{table*}

\section{Results}
\label{s:Results}
\subsection{Metrics}
\label{ss:5Metrics}
To determine performance between the experimental combinations, a total of  we use four main metrics: consistency of the filters (Section \ref{ss:5Consistency}), accuracy of vehicle state estimation (Section \ref{ss:5VehicleLocal}), accuracy of the final map estimation (Section \ref{ss:5LandmarkLocal}), and number of landmarks initialized (Section \ref{ss:5NumLM}).

The consistency of each algorithm and sensing strategy is evaluated using the normalized estimation error squared (NEES) at each time step $k$:
\begin{equation}
    \label{eqn:nees}
    \epsilon_k = (\hat{\mathbf{x}}_k - \mathbf{x}_k)^\textrm{T}P_k^{-1}((\hat{\mathbf{x}}_k - \mathbf{x}_k)).
\end{equation}
\noindent If the filter is consistent, then the state error will have an expected value of zero $E[\hat{\mathbf{x}}_k - \mathbf{x}_k] = 0$ and the true error covariance will match the covariance calculated by the filter $E[(\hat{\mathbf{x}}_k - \mathbf{x}_k)(\hat{\mathbf{x}}_k - \mathbf{x}_k)^\textrm{T}] = P_k$. Knowing $\epsilon_k$ is the sum of squares of independent, standard normal variables, if the above is true, $\epsilon_k$ will be sampled from a chi-square ($\chi^2$) distribution with an expected value equal to the number of states of the vehicle pose $E[\epsilon_k] = N_\mathbf{x} = 3$ \cite{bar2004estimation}. We average over the number of iterations and consider the average NEES (ANEES):
\begin{equation}
    \label{eqn:anees}
    \bar{\epsilon}_k = \frac{1}{N_{MC}}\sum_{i=1}^{N_{MC}}\epsilon_{i_k}.
\end{equation}

\noindent Here, the value $N_{MC}\bar{\epsilon}_k$ is $\chi^2$ distributed with $N_{MC}N_x$ degrees of freedom. A $1 - C = 95\%$ probability concentration region for this distribution $[\epsilon_1,\epsilon_2]$ can be found using:
\begin{equation}
    \label{eqn:neesConfidence}
    \begin{matrix}
        \text{$\epsilon_1 = \frac{\chi_{N_{MC}N_x}^2(C/2)}{N_{MC}}$}; \text{and }
        \text{$\epsilon_2 = \frac{\chi_{N_{MC}N_x}^2(1-C/2)}{N_{MC}}$}
    \end{matrix}.
\end{equation}

\noindent When the ANEES is in this region, the filter can be considered to be consistent but, if it is higher or lower than the bounds of this region, the filter is considered over confident or under confident respectively. 

In this work, ANEES is also used to determine if an iteration of a filter has diverged and is therefore not meaningful for analysis. For the EKF-SLAM algorithm, a maximum value of $\max_k \{\epsilon_k\} = 50$ was chosen which excluded 8 of the 115 Monte-Carlo iterations, while a maximum value of $\max_k \{\epsilon_k\} = 2,750$ was chosen for FastSLAM 2.0 which excluded 15 iterations. This value may seem extreme in the second case, but it is well known that FastSLAM 2.0 quickly becomes more over confident than EKF-SLAM and that the ANEES for both algorithms continues to rise over time \cite{bailey2006EKFconsistency,bailey2006Fastconsistency}.

The accuracy of the vehicle localization is analyzed using the root mean square error (RMSE) of the true model compared to the estimated position for each timestep $k$ and averaged over the length of the simulations $N_T$:
\begin{equation}
    \label{eqn:vehicleRMSE}
    \textrm{RMSE}_v = \sqrt{\frac{\sum_{k=1}^{N_T} (\hat{x}_{v_k} - x_{v_k})^2 + (\hat{y}_{v_k} - y_{v_k})^2}{N_T}}
\end{equation}
\noindent This same value is calculated for the error in the robot's bearing $(\hat{\theta}_{v_k} - \theta_{v_k})^2$ and is presented in appendix A.

The accuracy of the map building for both algorithms is analyzed in a similar way, by estimating the RMSE of the true landmark positions compared to the estimated positions of fully initialized landmarks, averaged over the number of fully realized landmarks $N_{mf}$:
\begin{equation}
    \label{eqn:mapRMSE}
    \textrm{RMSE}_m = \sqrt{\frac{\sum_{i=1}^{N_{mf}} [(\hat{x}_{i} - x_{i})^2 + (\hat{y}_{i} - y_{i})^2]}{N_{mf}}}
\end{equation}
\noindent While calculating RMSE, only fully initialized landmarks are considered, as the comparison must be made to some true value.

Finally, to help evaluate how effectively the vehicle can scan its surroundings, the number of landmarks that the vehicle is able to fully and partially initialize is analyzed.

We note that, as the directivity, and therefore HPBW, is determined by the simulated radius of the emitter (\ref{eqn:pistonModel}), the displayed emitter HPBWs in the result figures are rounded approximations and not exact values. This makes the data presentation more clear. Exact values can be found in appendix B.

\subsection{Filter Consistency}
\label{ss:5Consistency}
\begin{figure}
    \centering
    \includegraphics{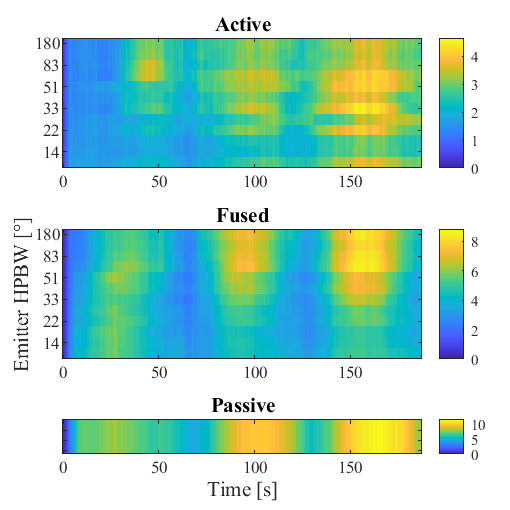}
    \caption{ Normalized estimation error squared (NEES) for EKF-SLAM averaged over 107 Monte-Carlo iterations.}
    \label{fig:neesEKF}
\end{figure}
\begin{figure}
    \centering
    \includegraphics{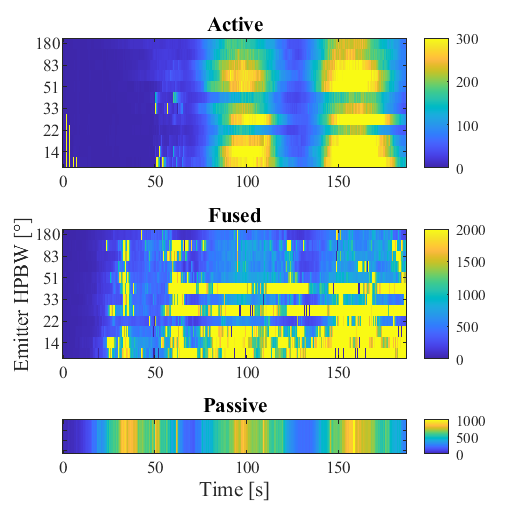}
    \caption{Normalized estimation error squared (NEES) for FastSLAM 2.0 averaged over 100 Monte-Carlo iterations.}
    \label{fig:neesFast}
\end{figure}

Figs. \ref{fig:neesEKF} and \ref{fig:neesFast}, show the average NEES (ANEES) for all Monte-Carlo iterations for EKF-SLAM and FastSLAM 2.0 respectively. It is immediately apparent from the magnitude attained by the ANEES that the EKF-SLAM algorithm is much more consistent than the FastSLAM 2.0 algorithm. Also, we notice a periodic pattern in the amplitude of the ANEES for all experimental combinations and  a general increase in maximum amplitude over time. The periodicity coincides with the noisy circular vehicle path, with low ANEES occurring when the vehicle is close to its starting position. These continuous loop closures reduce uncertainty in the entire map and lead to the general increase in maximum ANEES we observe as the vehicle is far from its starting position.

With $107$ Monte-Carlo iterations usable according to the limit described in (\ref{eqn:neesConfidence}), we consider our EKF-SLAM algorithm consistent for NEES values in the interval $[2.55, 3.48]$. Looking at Fig. \ref{fig:neesEKF}, we see the active strategy stays closer to an ideal level of confidence for a longer period of time compared to the passive strategy which becomes more overconfident after the first loop closure. The fused strategy has aspects of both others, remaining more consistent for longer periods of time while also being more confident in general by the end of the simulation. A trend towards greater consistency is also apparent as the HPBW of the emitter decreases in the fused sensing case.

With $100$ Monte-Carlo iterations usable according to the limit described in (\ref{eqn:neesConfidence}), we consider the FastSLAM 2.0 algorithm consistent for NEES values in the interval $[2.54, 3.50]$. The FastSLAM 2.0 ANEES results in Fig. \ref{fig:neesFast} share some similarities with the EKF-SLAM results, such as the cyclical nature of the active and passive strategies and an increase in ANEES over time, but the confidence of the FastSLAM 2.0 algorithm is much higher, which is to be expected \cite{bailey2006EKFconsistency, bailey2006Fastconsistency}. Results for the ANEES of the fused sensing strategy seem to be anomalous, with maximum ANEES extending far past the maximum values of the other sensing strategies. We believe this is due to the bearing-only landmark initialization technique not originally being created for FastSLAM 2.0. A different weighting strategy for landmark hypotheses may need to be employed to further limit the magnitude of changes in robot pose uncertainty.

\subsection{Vehicle Localization Accuracy}
\label{ss:5VehicleLocal}
\begin{figure*}[ht]
    \centering
    \includegraphics{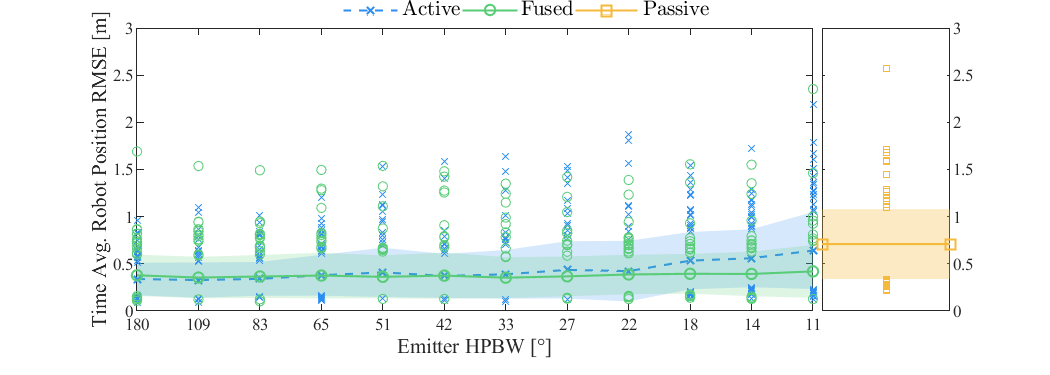}
    \caption{RMSE of vehicle position for EKF-SLAM averaged over the entire time of a single iteration. The marked lines represent the mean of the data, the highlighted section is one standard deviation from the mean, and all individual markers are iterations whose time averaged RMSE lie above said one standard deviation.}
    \label{fig:roboPoseEKF}
\end{figure*}
\begin{figure*}[ht]
    \centering
    \includegraphics{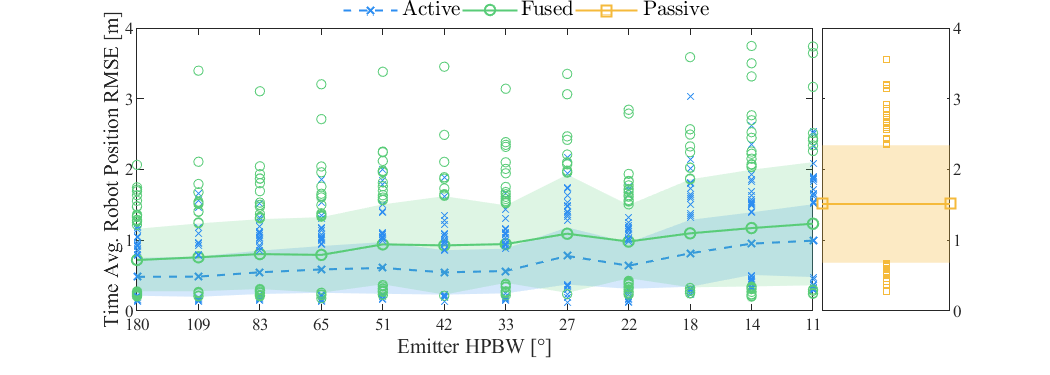}
    \caption{RMSE of vehicle position for FastSLAM 2.0 averaged over the entire time of a single iteration. The marked lines represent the mean of the data, the highlighted section is one standard deviation from the mean, and all individual markers are iterations whose time averaged RMSE lie above said one standard deviation.}
    \label{fig:roboPoseFast}
\end{figure*}

Figs. \ref{fig:roboPoseEKF} and \ref{fig:roboPoseFast} show the time averaged RMSE of the position of the vehicle in the map for EKF-SLAM and FastSLAM 2.0 respectively (bearing RMSE figures can be found in appendix A1). Both estimation algorithms follow the same general trends. Passive sensing has a greater mean and standard deviation in RMSE than both the active and fused strategies at any point, and the variance in performance of fused and passive sensing is generally greater than that of active sensing. Overall, EKF-SLAM has lower mean RMSE and variance than FastSLAM 2.0 for all experimental combinations. We also can see that, as HPBW decreases, the performance of fused sensing approaches the performance of passive sensing. This is to be expected, as the limit of the HPBW approaching zero for fused sensing is the passive sensing strategy

Notably in Fig. \ref{fig:roboPoseEKF}, for HPBWs less than approximately $42^\circ$, vehicle localization using fused sensing shows an improvement over active sensing on average, although this improvement never reaches statistical significance. This is opposed to FastSLAM 2.0 where fused sensing never improves upon the performance of active sensing when estimating robot position, though the wide variance of fused sensing RMSE means that the performance of active alone is not significantly better. 

\subsection{Landmark Localization Accuracy}
\label{ss:5LandmarkLocal}
\begin{figure}
    \centering
    \includegraphics{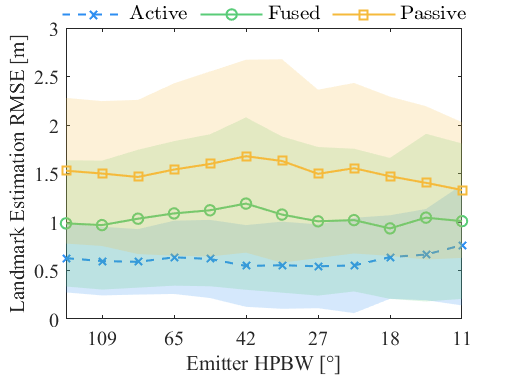}
    \caption{RMSE of final positions for EKF-SLAM where only landmarks seen by each sensing strategy were included in the calculation. The marked lines represent the mean of the data and the highlighted sections are $\pm$ one standard deviation from the mean.}
    \label{fig:featureEKF_same}
\end{figure}
\begin{figure}
    \centering
    \includegraphics{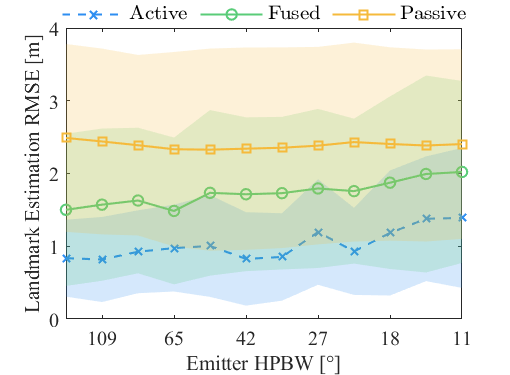}
    \caption{RMSE of final positions for FastSLAM 2.0 where only landmarks seen by each sensing strategy were included in the calculation. The marked lines represent the mean of the data and the highlighted sections are $\pm$ one standard deviation from the mean.}
    \label{fig:featureFast_same}
\end{figure}

To analyze the performance of the landmark position estimation, figs. \ref{fig:featureEKF_same} and \ref{fig:featureFast_same},representing EKF-SLAM and FastSLAM 2.0 respectively, show the RMSE of only the landmarks that were seen by all sensing strategies for a given Monte-Carlo iteration. While the average performance of fused sensing never outperforms that of the active strategy, it is within approximately one standard deviation of the mean most of the time. In these figures, we can also observe the expected dependence of fused sensing performance on both active and passive accuracy. We notice, as HPBW decreases, the fused sensing RMSE tends to follow the general trend of passive sensing while its magnitude is decreased by the RMSE of active sensing. Again, the RMSE of fused and passive estimations will approach each other as HPBW approaches zero due to the limit described previously.

\subsection{Number of landmarks initialized}
\label{ss:5NumLM}

\begin{table}[h!]
\begin{center}
\caption{Comparing Fully Initialized Landmarks to the Total Number of Sensed Landmarks ($\mu \pm 1\sigma$)}
\label{tab:FullvTotInit}
\begin{tabular}{| c | c | c | c | c |}
\hline
\textbf{HPBW}& \multicolumn{2}{c||}{\textbf{EKF-SLAM}} &\multicolumn{2}{c|}{\textbf{FastSLAM 2.0}}\\
\hline
 & Fully & \multicolumn{1}{c||}{Total} & Fully & Total\\
\hline
$180^\circ$& $37.7 \pm 2.9$& \multicolumn{1}{c||}{$43.7 \pm 2.6$}& $39.3 \pm 2.8$& $43.7 \pm 2.4$\\
\hline
$33^\circ$& $31.9 \pm 3.3$& \multicolumn{1}{c||}{$40.8 \pm 3.2$}& $32.6 \pm 3.3$& $40.7 \pm 3.1$\\
\hline
$11^\circ$& $30.0 \pm 3.3$& \multicolumn{1}{c||}{$40.1 \pm 3.2$}& $30.5 \pm 3.2$& $40.2 \pm 3.1$\\
\hline
Passive $(0^{\circ})$& $26.6 \pm 3.3$& \multicolumn{1}{c||}{$39.5 \pm 3.4$}& $22.1 \pm 4.9$& $39.4 \pm 3.3$\\
\hline
\end{tabular}
\end{center}
\end{table}

\begin{figure}
    \centering
    \includegraphics{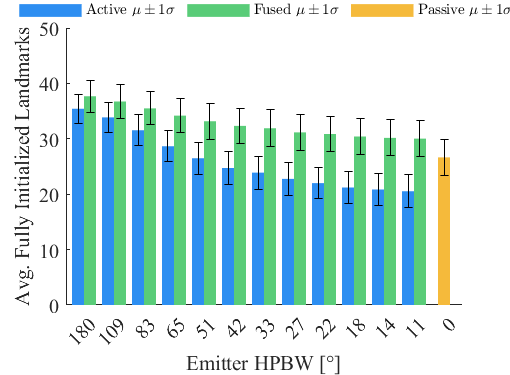}
    \caption{Average number of fully initialized landmarks for active, passive, and fused sensing with EKF-SLAM. Error bars are one standard deviation from the mean.}
    \label{fig:featureInitEKF}
\end{figure}
\begin{figure}
    \centering
    \includegraphics{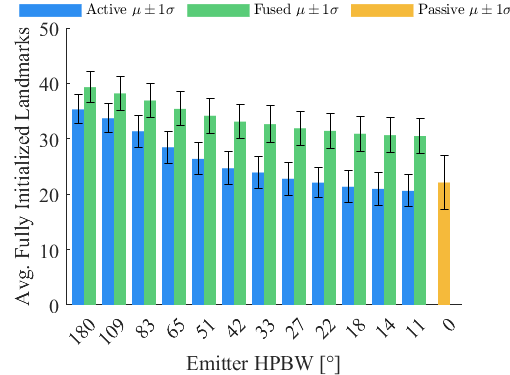}
    \caption{Average number of fully initialized landmarks for active, passive, and fused sensing with FastSLAM 2.0. Error bars are one standard deviation from the mean.}
    \label{fig:featureInitFast}
\end{figure}

Figs. \ref{fig:featureInitEKF} and \ref{fig:featureInitFast}, EKF- and FastSLAM 2.0 respectively, show the average number and standard deviation ($\mu \pm 1\sigma$) of fully initialized landmarks across all HPBWs and for passive sensing. Table \ref{tab:FullvTotInit} lists data for a selection of HPBWs for EKF- and FastSLAM 2.0 for both fully initialized and partially initialized landmarks. Actively initialized landmarks are not present in Table \ref{tab:FullvTotInit} as they are automatically fully initialized due to range and bearing measurements.

A clear trend towards fewer landmarks being initialized as the HPBW decreases can be seen for both SLAM algorithms when active sensing is used. When passive sensing is implemented, an increase in the number of fully initialized landmarks that is quite significant in the case of EKF-SLAM is present, whereas FastSLAM 2.0 has a standard deviation that is larger than any other experimental combination. We can also see that the total number of landmarks seen by passive sensing is greater than any time active sensing is used, and that the means differ by at least one standard deviation for HPBWs less than $65^\circ$. 

When both strategies are combined in the fused sensing case, influences of both strategies can be observed. The trend in fewer landmarks being initialized compared to active sensing is lessened but still present, and the difference in the number of fully initialized landmarks becomes more significant as the HPBW decreases. Looking at Table \ref{tab:FullvTotInit}, we also see a trend of decreasing average of fully and total initialized landmarks with an increase in the standard deviation. This is to be expected due to fused and passive sensing becoming equivalent when HPBW is zero.

\section{Discussion}
\label{s:Discuss}

Previous work investigating fused acoustic sensing found that it is mainly beneficial in terms of reduced error and uncertainty when active sensing is limited to narrow angles and in cases where bearing measurement noise is low ($\sigma_\phi = [0.1,0.2]$ rad) \cite{jahromi2021eavesdropping}. Our results cannot be exactly compared to this work, as it assumed a standard deviation of range noise $\sigma_d = 0.01$ m, while our simulated sensor system has a larger value of $\sigma_d = 0.2$ m, but our results are consistent with the findings, at least when comparing the EKF-SLAM results. 

We have found that EKF results are better than the results from FastSLAM 2.0 in almost every category except for the number of fully and partially initialized landmarks, where results are very similar. As stated previously, FastSLAM 2.0 is known to quickly become overconfident and the continuous loop closures experienced in this system lead to a loss in particle diversity that heavily reduces performance. Further, the bearing-only landmark technique used greatly reduces the speed of each prediction and update due to the multiple weighting calculations introduced for each landmark hypothesis. This quick degradation along with a relatively inaccurate sensor may be the cause of the relatively poor performance we observe.

Using an EKF algorithm seems to be the most promising for future work, as fused sensing has at least comparable if not better performance to active sensing when looking at number of landmarks fully initialized and the localization of the robot in its map. Even considering the accuracy of the map generated by the robot presented in fig. \ref{fig:featureEKF_same}, we see that while not as accurate on average, the standard deviations of fused and active sensing begin to overlap more as the emitter HPBW decreases. This leads us to a promising conclusion that a more accurate sensor could further improve the performance of fused sensing to at least match the abilities of active sensing when localizing landmarks.

\section{Conclusion}
\label{s:Conclude}
In summary, our simulations are consistent with previous work that fused sensing can prove just as useful or better than active sensing in low noise environments. We also found that a system using a strictly FastSLAM 2.0 approach may not be very promising due to its difficulty in implementation and errors in estimation that are greater than those of the more straightforward EKF-SLAM. The performance of fused sensing when data association is not known and the static beacon is a moving vehicle itself is still of interest. This future direction comes with its own challenges but this paper continues to support an interesting path of research into how acoustic sensing may be used for collaborative navigation in teams of robots. 

\section*{Acknowledgments}

This work was funded by the National Science Foundation under the Graduate Research Fellowship Program to A.J.B. and under grant 1751498.

{\appendices

\section{Time Averaged Robot Bearing RMSE}
\begin{figure*}
    \centering
    \includegraphics{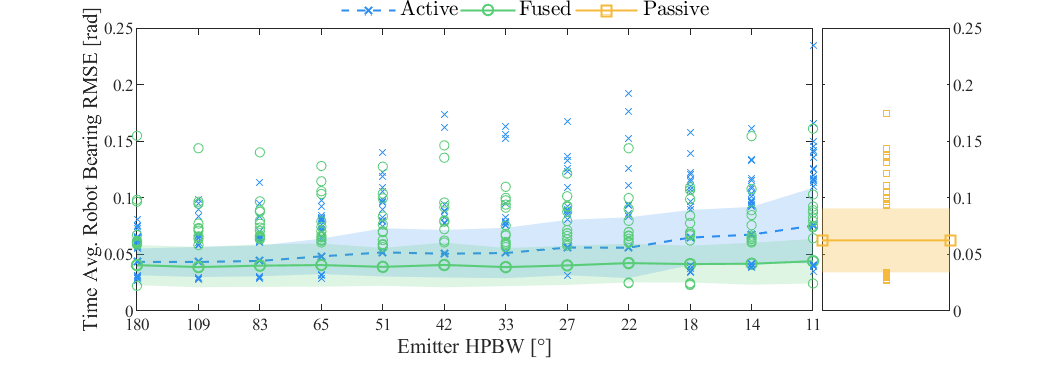}
    \caption{EKF-SLAM: RMSE of vehicle bearing averaged over the entire time of a single iteration. The marked lines represent the mean of the data, the highlighted section is one standard deviation from the mean, and all individual markers are iterations whose time averaged RMSE lie above said one standard deviation.}
    \label{fig:roboBearingEKF}
\end{figure*}
\begin{figure*}
    \centering
    \includegraphics{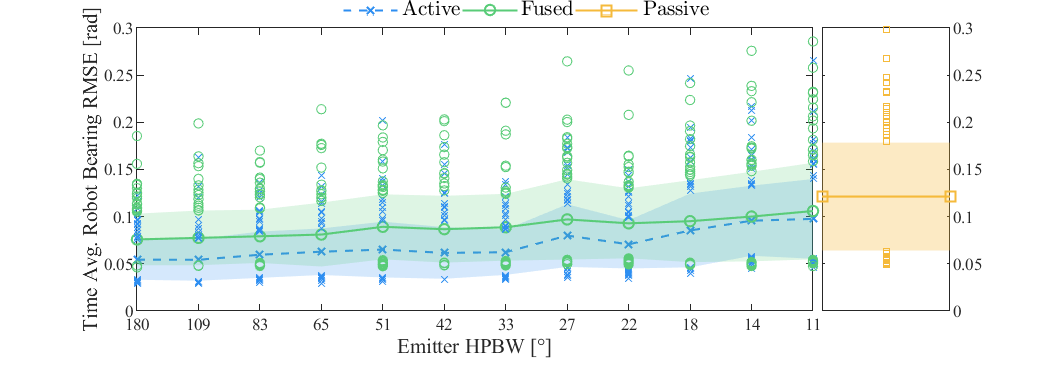}
    \caption{FastSLAM 2.0: RMSE of vehicle bearing averaged over the entire time of a single iteration. The marked lines represent the mean of the data, the highlighted section is one standard deviation from the mean, and all individual markers are iterations whose time averaged RMSE lie above said one standard deviation.}
    \label{fig:roboBearingFast}
\end{figure*}

In Figs. \ref{fig:roboBearingEKF} and \ref{fig:roboBearingFast}, we see many of the same trends discussed in Section V.C of the main body of the paper. EKF-SLAM is again more accurate when localizing the vehicle for all strategies and HPBWs. When using EKF-SLAM, the average RMSE of bearing estimation for fused sensing is less than active only for all HPBWs. This is compared to the fused sensing strategy only becoming more accurate after a HPBW of around $42^\circ$ when estimating position.

\section{List of Precise Half-Power Bandwidths}
In all results figures, the HPBWs of the emitter have been rounded to the nearest whole number to improve data display and understanding. As the values are based on the simulated radius of the emitter \ref{eqn:pistonModel}, which are logarithmically spread between the values of $[2.5,25]$ mm, when converted to degree values they may seem arbitrarily chosen. Here is a more precise list of the values in degrees: $[11.4, 14.2, 17.6, 21.8, 26.8, 33.2, 41.6, 51.4, 64.8, 83,\\ 108.6, 180]$.
}


\newpage

\section{Biographies}
\begin{IEEEbiographynophoto}{Aidan J. Bradley}
is a fourth year Ph.D. student in the Engineering Mechanics programs at Virginia Tech. He received his B.S. in Engineering with a specialization in Electrical Engineering from Roger Williams University (2018). In 2020, Aidan was awarded an NSF Graduate Research Fellowship to investigate the control of multi-agent robotic teams using bioinspired sensing capabilities.
\end{IEEEbiographynophoto}

\begin{IEEEbiographynophoto} 
{Nicole Abaid} received her Ph.d. in mechanical engineering from the Polytechnic Institute of New York University in 2012. She is currently an associate professor in the Department of Mathematics at Virginia Tech. Her research interests include biologically-inspired robotics and multi-agent systems. 
\end{IEEEbiographynophoto}

\vfill

\end{document}